\documentclass[twocolumn]{aastex631}

\begin{document}

\title{Binary fraction in Galactic star clusters: FSR 866, NGC 1960, and STOCK 2}


\author{\firstname{L.}~\surname{Yalyalieva}}
\affiliation{Physics Department, Lomonosov Moscow State University, Leninskie Gory, Moscow, 119991, Russian Federation}
\affiliation{Sternberg Astronomical Institute, Lomonosov Moscow State University, Universitetsky pr.13, Moscow, 119234, Russian Federation}

\author{\firstname{A.}~\surname{Chemel}}
\affiliation{Physics Department, Lomonosov Moscow State University, Leninskie Gory, Moscow, 119991, Russian Federation}
\affiliation{Sternberg Astronomical Institute, Lomonosov Moscow State University, Universitetsky pr.13, Moscow, 119234, Russian Federation}

\author{\firstname{G.}~\surname{Carraro}}
\email{giovanni.carraro@unipd.it}
\affiliation{Department of Physics and Astronomy, Padova University, Vicolo Osservatorio 3, I-35122, Padova, Italy}

\author{\firstname{E.}~\surname{Glushkova}}
\affiliation{Physics Department, Lomonosov Moscow State University, Leninskie Gory, Moscow, 119991, Russian Federation}
\affiliation{Sternberg Astronomical Institute, Lomonosov Moscow State University, Universitetsky pr.13, Moscow, 119234, Russian Federation}



\begin{abstract}

The study of binary stars in different astronomical environments offers insights into the dynamical state of the hosting stellar systems. The Binary Fraction in fact plays a crucial role in the dynamical evolution of stellar system, regulating processes like  mass segregation and dynamical heating, and in some cases leading to the formation exotic object, like for instance blue straggler stars. We used two methodologies to estimate the binary fraction in three different-age open star clusters: FSR 866, NGC 1960 (M36), and Stock 2. The first, a photometric approach based on colour-magnitude diagram analysis, and the second, a spectroscopic technique which employs radial velocity measurements. We used Gaia DR3 data in tandem with new spectroscopic observations, and employed the DBSCAN clustering algorithm to identify probable cluster members based on proper motion and parallax in 3D space.
The new sample of cluster members allows us to provide new estimates of the cluster fundamental parameters. As a by-product, we found two previously undetected, small physical groups of stars in the background of NGC 1960.
The resulting binary fractions lie in the range 0.3 - 0.5 and are in good agreement with those expected theoretically for open clusters. 

\end{abstract}

\keywords{Open star clusters (1160) --- Binary stars (154)}


\section{Introduction} \label{sec:intro}

The fraction of binary stars exhibits variations in different stellar environments, such as open clusters, globular clusters and the Galactic field, as indicated by \citet{Borodina}. The binary fraction is an important parameter that affects the dynamical evolution of the parent cluster, causing a series of effects, from the dynamical heating and mass segregation to the creation of exotic systems. In the general Galactic disk field, however, the value of binary fraction appears to be connected to the binary fraction observed in open clusters, as established in a series of papers by \citet{Kroupa1995a, Kroupa1995b, Kroupa1995c}.

The binary fraction in clusters depends on various circumstances. \citet{MarksOh2011} demonstrated that denser clusters initially have lower binary fractions compared to looser systems of equivalent age. 
The same study also highlighted that in denser clusters more binaries are dissolved within a given timescale due to the higher collision rate. Moreover massive binary stars can be ejected from the parent clusters and the fraction of  binaries ejected strongly depends on the initial cluster density \citep{Oh2014, Oh2016}.

There are two main methods to estimate binary fraction in star clusters. The first one is  the photometric method based on identifying the binary status according to their location in the color-magnitude diagram of the parent cluster. This has been used in a large number of works, such as \citet{Khalaj, Sollima, Cordoni18}. The second one is based on spectroscopy and exploits time-series observations and the extraction and analysing of radial velocities. An example of such method is in \citet{Geller}, where the authors summarised their 45 years of observations.

In this study, we combined these two approaches to investigate three different age open star clusters and determine their binary fraction. 

The first cluster is FSR 866, an old OC in Gemini. Nonetheless, there are ambiguities in the estimates of the basic parameters of this cluster in the literature, see Table \ref{tab:Literature}.
The second cluster is the well known, relatively young cluster NGC 1960 (M~36). This OC is an interesting object for planet formation investigation, because all estimates of its age crowds in the 10 -- 30 Myr range, where rocky planets are likely to achieve their final masses according to up-to-date models of terrestrial planet formation \citep{SmithJeffries2012}. M 36 has been also the target for search and characterization of circumstellar discs \citep{Haisch2001}.
The third OC we studied is Stock 2, see the comparison of its parameters from the literature in the Table \ref{tab:Literature}.

\begin{table*}[h]
\caption{Mean values of the clusters' parameters culled from the literature }
\label{tab:Literature}
\centering
\begin{tabular}{lcllll}
\hline
Cluster  & $\log{(\mathrm{Age/yr})}$                   & Distance,                      & $V_r$, & Reference                   & Remarks\\
         &  & pc                             & km s$^{-1}$      &                             &\\
\hline
FSR 866  &9.57                    & 1215                           & 65.55            &\citet{Dias}                 & Based on 2 stars\\
         &9.31                    & 1211                           & 65.46            & \citet{Tarricq2021}         & Based on 2 stars\\
         &9.26                    & 1503                           & 52.04            & \citet{Zhong2020}           & \\
         &9.33                    & 1250                           &                  & \citet{CantatGaudin2020}    & \\
         &9.60                    & $\approx$ 1245                 &                  & \citet{He2022}              & Distance is approximately derived from \\
         &                        &                                &                  &                             & original parallax = 0.803 mas \\
\hline		 
NGC 1960 &7.20                    & 1300                           &                  & \citet{Sanner2000}          & 		\\
         &7.48                    & 1086                           &                  & \citet{Dias}                & Based on 285 potential cluster members\\
         &                        &                                & 7.03             & \citet{Tarricq2021}         & Based on 1 star\\
         &7.45                    & 1162                           &                  & \citet{CantatGaudin2020}    & Derived from Gaia DR2 data, on 280 stars\\
         &                       &                                 &                  &                             & with membership probability $\ge$ 70\%\\
         &7.565                   & 1200                           &-12.09            & \citet{Zhong2020}           & Uses LAMOST catalogue and membership data\\
         &                        &                                  &                  &                           & from \citet{CantatGaudin}\\
         &                        &                                &                  & \citet{Thompson2021}        & new binary stars detection technique\tablenotemark{a};\\
         &                        &                                &                  &                             & binary fraction = 0.66\\
         &7.44                    & 1170                           &                  & \citet{Joshi2020}           & A comprehensive long-term study of on the\\
         &                   &                           &                  &                                       & basis of Gaia DR2 kinematic data\\			 
         &                        &                                &                  & \citet{Seleznev2016}        & The structural parameters of seven OCs,\\
         &                        &                                &                  &                             & were derived with a Kernel Density Estimation\\
         &                        &                                &                  &                             & technique and N-body simulations\\
         &7.34                    &                                &                  & \citet{Jeffries2013J}       & Age estimation, based on age-dependent \\
         &                        &                                &                  &                             & position of "lithium depletion boundary"\\
         &7.3                     &                                &                  & \citet{Bell2013}            & Age estimate, derived from the fitting\\
         &                        &                                &                  &                             & of pre-main-sequence isochrones\\
         &7.30                    & $\approx$ 1180                    &               & \citet{He2022})             & Basing on 360 cluster members; distance is \\
         &                        &                      &                  &                                       & roughly derived from original parallax = 0.845 mas\\
\hline		 
Stock 2	 &8.905                   &374                             &8.757             & \citet{Dias}                & Based on 1157 members,\\
  &                    &                             &              &                                               & 185 radial velocity measurements\\
		 &                        &                                &8.61              & \citet{Tarricq2021}         & Based on $V_r$ measurements from Gaia-RVS \\
		 &                        &                                &8.61              &                             & data and ground-based surveys, 177 stars\\
		 &8.65                    &                                &8.0               & \citet{AlonsoSantiago2021}  & A spectroscopic study based on high-resolution \\
		 &                        &                                &                  &                             & spectroscopy from HARPS-N\\
		 &8.44                    &400                             &1.55              & \citet{Zhong2020}           & Use LAMOST data\\
		 &8.60                    &399                             &                  & \citet{CantatGaudin2020}    & Based on 1178 stars with membership \\
		 &                          &                              &                  &                             & probability higher than 50\%\\
		 &8.60                    &375                             &                  & \citet{Casamiquela2022}     & Aimed at searching for chemically peculiar stars\\
		 &                        &                                &7.08              & \citet{Carrera2019}          & Used radial velocity data from APOGEE\\
		 &                        &                                &-17.39            & \citet{Mermilliod2008}      & Based on one red giant, binary fraction = 0.67\\
		 &8.45                    & $\approx$ 374                  &                  & \citet{He2022}              & Based on 1471 probable members; distance is\\
		 &                        &                                               &                  &              &  roughly derived from original parallax = 2.674 mas\\
\hline
\end{tabular}
\tablenotetext{a}{the technique is applicable even for faint systems and based on comparison of observed in multiple filters magnitudes and synthetic star spectral energy distributions}
\end{table*}

As can be immediately realized from this overview, there is a general lack of spectral data in the literature for all three clusters, and for this reason the average radial velocities exhibit quite a significant scatter. Although the OCs photometric parameters derived by different authors show a somewhat better agreement, we decided in this study to re-determine them in order to then use homogeneous data to estimate the fraction of binary stars. To this aim,  the first piece of information we need to obtain is the cluster membership.

\section{CLUSTERING ALGORITHM}
\label{sec:clus_algo}
    
Cluster members were identified using cluster analysis methods.  Clustering was performed in 3D space of proper motions and parallaxes from Gaia DR3 \citet{GaiaDR3}. Only data with parallax errors smaller than 20 per\,cent were taken into account. We used the same technique described in \citet{ngc225}, namely a Python implementation of the Density-Based Spatial Clustering of Applications with Noise (DBSCAN) algorithm provided by the \textsc{scikit-learn} library \citep{scikit-learn}. DBSCAN requires two main parameters -- {\it  eps}, the maximum distance between two points for one of them to be considered a neighbour of the other, and {\it min\_samples}, the minimum number of points in the neighbourhood of a particular point for this to be considered a core point.
Before clustering, the coordinates were scaled to unit dispersion and a principal component analysis was performed to exclude possible dependencies between coordinates. This approach also leads to the parameter {\it  eps} being dimensionless.
We run DBSCAN  for a set of {\it  eps} and {\it min\_samples} parameters. The range for the {\it  eps} parameter for all clusters was [0.01, 0.99].

The testing limits {\it min\_samples} were found for each of these three open clusters separately as the range where the result of the clustering provides two groups - the cluster, as a group of physically connected stars, and the background stars.

The star cluster group was identified by comparing the mean values of the group proper motion and parallaxes with the corresponding values as listed in \citet{Dias}.
The limits for {\it min\_samples} depend on the cluster size and dispersion of its parameter and for our open clusters they turned out to be as follows: FSR866 - [1, 350], NGC1960 - [1, 750], Stock2 - [1, 2050].
The steps of varying parameter {\it  eps} is 0.01 and for {\it min\_samples} is 1. However, due to the large size and therefore large number of objects in the case of Stock2, the step in {\it  eps} was increased to 0.03 and in {\it min\_samples} to 20.

When all N runs that satisfy the criterion were completed (those that give exactly two groups of stars), the probability of each star to be an open cluster member is derived as the number of times when the star was labeled as a cluster member among these N runs divided by N itself.

Finally, we collect all the stars designated as cluster members based on membership probability in descending order of $G_{\rm{mag}}$-magnitude from Gaia DR3 and assigned indices to them according to that order.

\begin{figure}[h]
\plotone{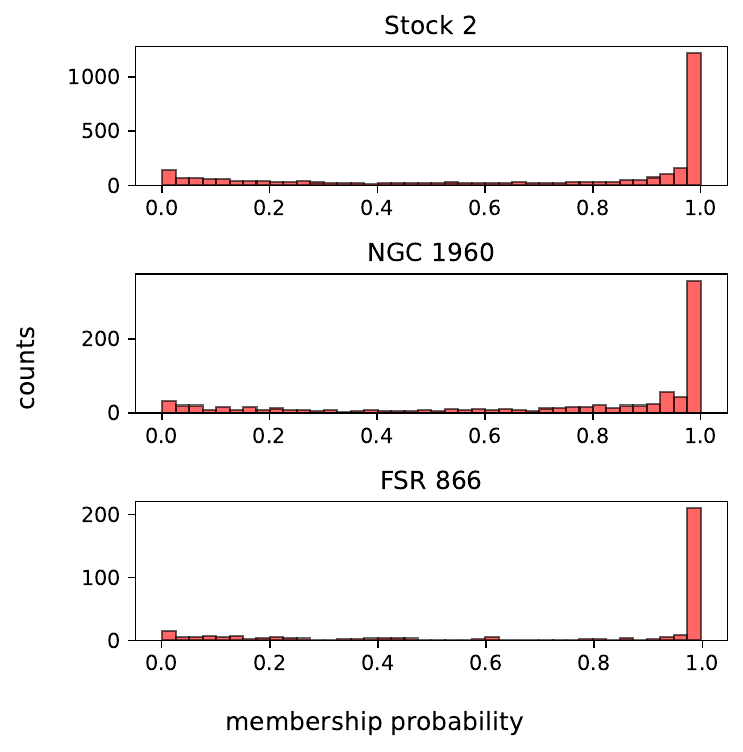}
\caption{Membership probability distribution for stars in the field of Stock 2, NGC 1960 and FSR 866. The final threshold probability level was chosen manually in the area where a sharp increase in the number of members begins. For Stock 2 and NGC 1960, this threshold is 0.95, and  0.985 for FSR 866.}
\label{fig:hist_prob}
\end{figure}

The distributions of membership probabilities for Stock 2, NGC 1960, and FSR 866 are shown in Figure \ref{fig:hist_prob}. Despite the strong peaks close to the unity, the total number of cluster members amongst all the stars are comparable for all clusters with the number of field stars: about 49\% member-stars for Stock 2 ($P \ge 0.95$), 45\% for NGC 1960 ($P \ge 0.95$) and 62\% for FSR 866 ($P \ge 0.985$).

\begin{figure*}[h]
\plotone{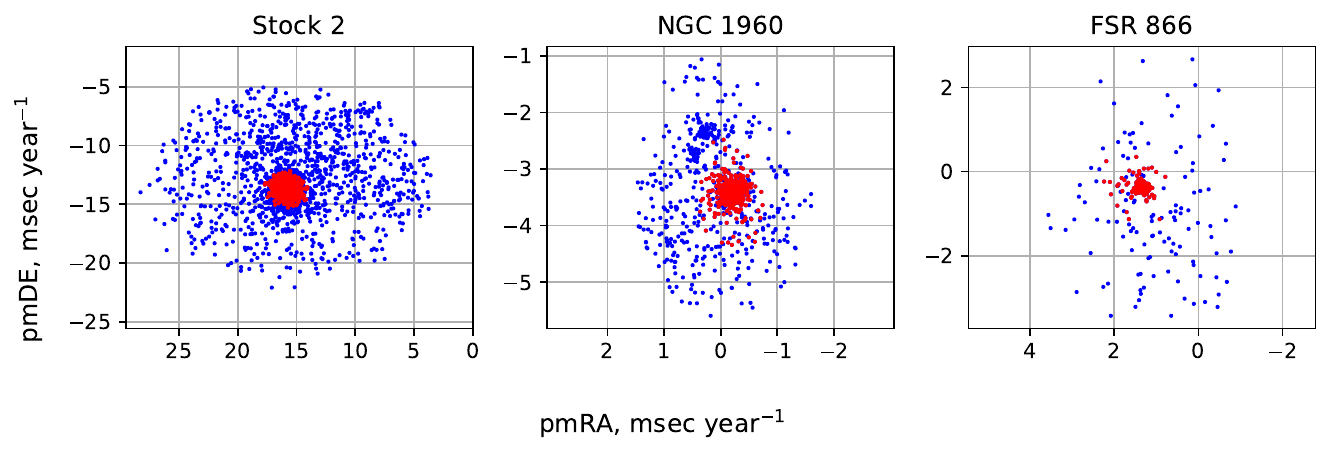}
\caption{Proper motion distribution for cluster members (red circles) and field stars (blue dots) in the field of Stock 2, NGC 1960 and FSR 866, derived on the basis of membership probability. The probability threshold for membership is 0.95, 0.95, and 0.985, respectively. Proper motions are from Gaia DR3. On the second panel, that corresponds to NGC 1960, two  concentrations besides the cluster are visible, and are discussed in Sec. \ref{sec:clus_algo}}
\label{fig:pm_distr}
\end{figure*}

Figure \ref{fig:pm_distr} shows the position of probable members (red circles) and field stars (blue dots) in the proper motion space, where members occupy a much smaller area, as expected.

\begin{figure}[h]
\plotone{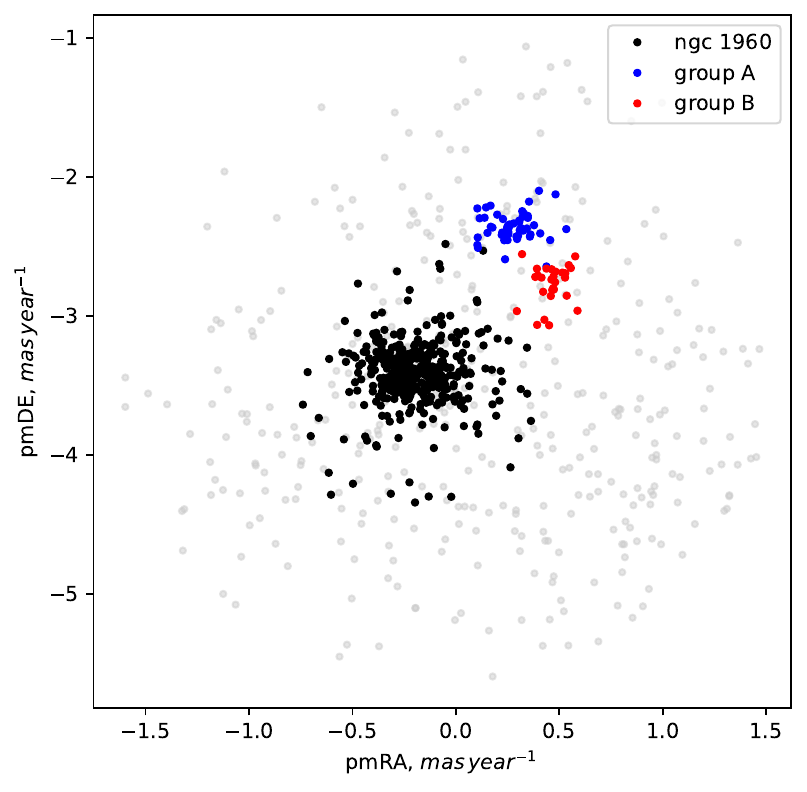}
\caption{Stars in the field of NGC 1960 in the proper motion space. Member-stars of NGC 1960 itself are depicted with black dots, while stars of groups A and B with blue and red dots, respectively.}
\label{fig:ABgr_pm}
\end{figure}

\begin{figure}[h]
\plotone{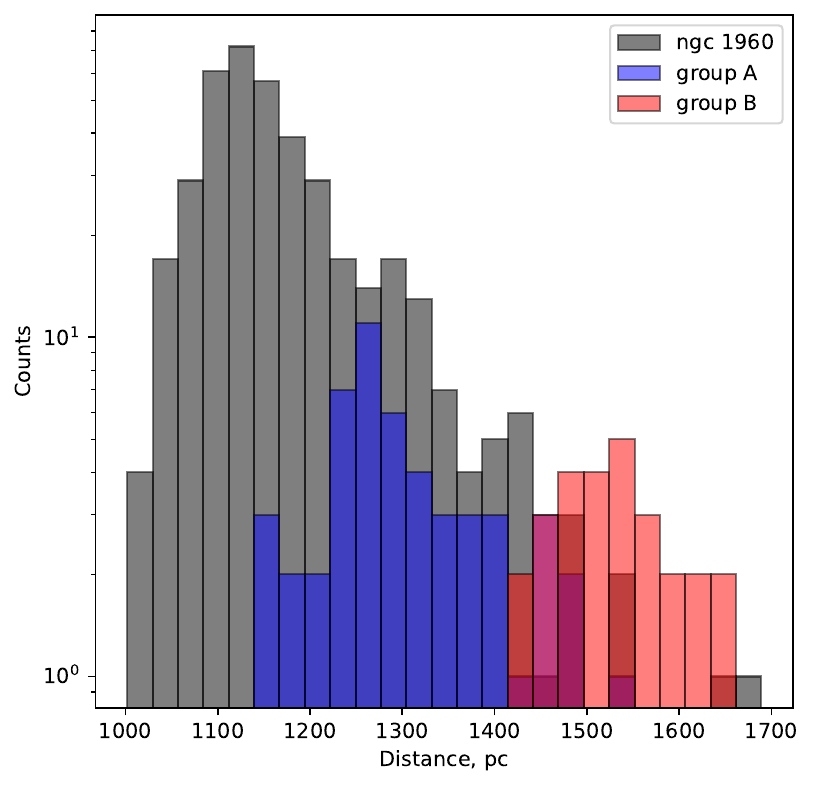}
\caption{Distance distribution for stars in the field of NGC 1960. Colors are the same as in Fig. \ref{fig:ABgr_pm}. Based on geometric distances from \citet{BailerJones}.}
\label{fig:ABgr_bj}
\end{figure}

\begin{figure}[h]
\plotone{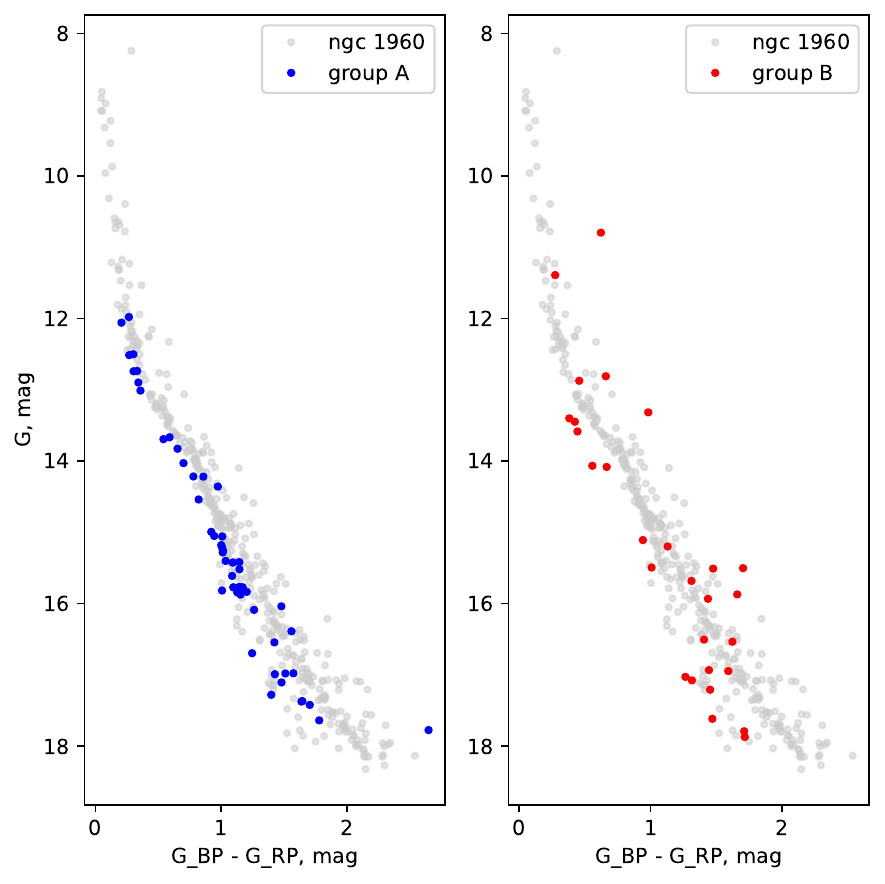}
\caption{Color-magnitude diagram of two studied groups in comparison with CMD of NGC 1960 (gray dots). Colors for two groups are the same as in Fig. \ref{fig:ABgr_pm}.}
\label{fig:ABgr_cmd}
\end{figure}

As one can readily see, in the vector point diagram for stars in the direction of the NGC 1960 cluster, in addition to the stars of the cluster, two concentrations are clearly distinguished: group A around (0.25, -2.3) mas~yr$^{-1}$ and group B at about (0.25, -2.7) mas~yr$^{-1}$. Using the same clustering method as for three studied open clusters, we identified the stars belonging to these groups. We proceeded as follows. First, we excluded from the sample the stars in the NGC 1960 field, whose membership probability is more than 95\%. We considered then clustering options for the remaining stars in the space of proper motions and parallaxes for clustering parameters 0 $<$ \emph{eps} $<$ 1 with a step of 0.01,
1 $<$ \emph{min\_samples} $<$ 100 with a step of 1. We chose those parameters for which 3 groups are obtained as a result of the clustering algorithm: groups A and B, and a group of background stars. We left the stars, which were labeled as belonging to groups in at least half of the cases and obtained a list of 51 member-stars for group A and of 28 for group B (Tables \ref{tab:groupA} and \ref{tab:groupB}, respectively). Figures \ref{fig:ABgr_pm}, \ref{fig:ABgr_bj} and \ref{fig:ABgr_cmd} show the positions of member-stars of the two groups in proper motion space, in distances and in the color-magnitude diagram in comparison with NGC 1960 stars. Table \ref{tab:groups_mean} lists mean proper motions and mean geometric distance from \citep{BailerJones} of the groups A and B. As it can be noticed, the objects for these groups have, on average, larger distance than NGC 1960. The color-magnitude diagrams overlap with the one of NGC 1960, but indicate a slightly larger shift in stellar magnitude, which corresponds to a slightly larger distance modulus. We did not find any reference about these groups in the literature. 

\begin{table}[h]
\caption{Gaia DR3 Source IDs for member-stars in group A}
\label{tab:groupA}
\centering
\begin{tabular}{cc}
\hline
\multicolumn{2}{c}{Group A, Gaia DR3 Source ID}\\
\hline
3449535625534281728 & 3449540337117675904 \\
3449537584039369216 & 3449545662877131264 \\
3449620219212841856 & 3449621593602450304 \\
3449537760137369984 & 3449538649192368256 \\
3449541844646924160 & 3449532640536358400 \\
3449525047034490752 & 3449527830173281920 \\
3449538065075704448 & 3449582874473917696 \\
3449538030715966848 & 3449519682618909952 \\
3449525081394227712 & 3449308438947124480 \\
3449534599041515776 & 3449537828856836224 \\
3449621494819206784 & 3449513845759836160 \\
3449537931936045696 & 3449542394404128000 \\
3449635925909191424 & 3449537622698406784 \\
3449525356272153856 & 3449526661942171776 \\
3449297516845882752 & 3449534599041515008 \\
3449535732912829184 & 3449525837304812288 \\
3449558032383111552 & 3449540642057682432 \\
3449537622698406272 & 3449526593222697600 \\
3449620425371207936 & 3449583767827067264 \\
3449621357380259456 & 3449294660691797632 \\
3449321018906842624 & 3449538030717064064 \\
3449427461081258752 & 3449539576905628416 \\
3449537897576312064 & 3449527516639376000 \\
3449622078934767360 & 3449524561701743360 \\
3449524943955265920 & 3449588573892846336 \\
3449306725255687808 & \\
\hline
\end{tabular}
\end{table}

\begin{table}[h]
\caption{Gaia DR3 Source IDs for member-stars in group B}
\label{tab:groupB}
\centering
\begin{tabular}{cc}
\hline
\multicolumn{2}{c}{Group B, Gaia DR3 Source ID}\\
\hline
3449502335247586432 & 3449625068231967616 \\
3449511440578192256 & 3449511612376874112 \\
3449411793037536640 & 3449588578190417024 \\
3449316483421323776 & 3449413584041915648 \\
3449428423152996864 & 3449524394195803520 \\
3449424166841367296 & 3449519407740951424 \\
3449427396656719488 & 3449315693147337728 \\
3449427293577508224 & 3449518656123440384 \\
3449314078239645184 & 3449299711573599744 \\
3449431176227952256 & 3449317926530426368 \\
3449325726190988800 & 3449520541612289536 \\
3449619742472525184 & 3449588956147535360 \\
3449531021330051712 & 3449321564367376000 \\
3449436325893248256 & 3449514597377360000 \\
\hline
\end{tabular}
\end{table}

\begin{table}[h]
\caption{Mean proper motion, geometric distance and its standard deviation for groups A and B. Distances are from \citep{BailerJones}}
\label{tab:groups_mean}
\centering
\begin{tabular}{lcccc}
\hline
Group & $<$pmRA$>$ & $<$pmDE$>$ & $<$rgeo$>$ & std(rgeo)\\
& mas yr$^{-1}$ & mas yr$^{-1}$ & pc & pc \\
\hline
A & 0.275 & $-$2.358 & 1302.5 & 91.5 \\
B & 0.462 & $-$2.771 & 1536.3 & 69.0 \\
\hline
\end{tabular}
\end{table}

\section{PHOTOMETRIC DISTANCE AND AGE}
\label{sec:photDist}
\begin{figure*}[h]
\plotone{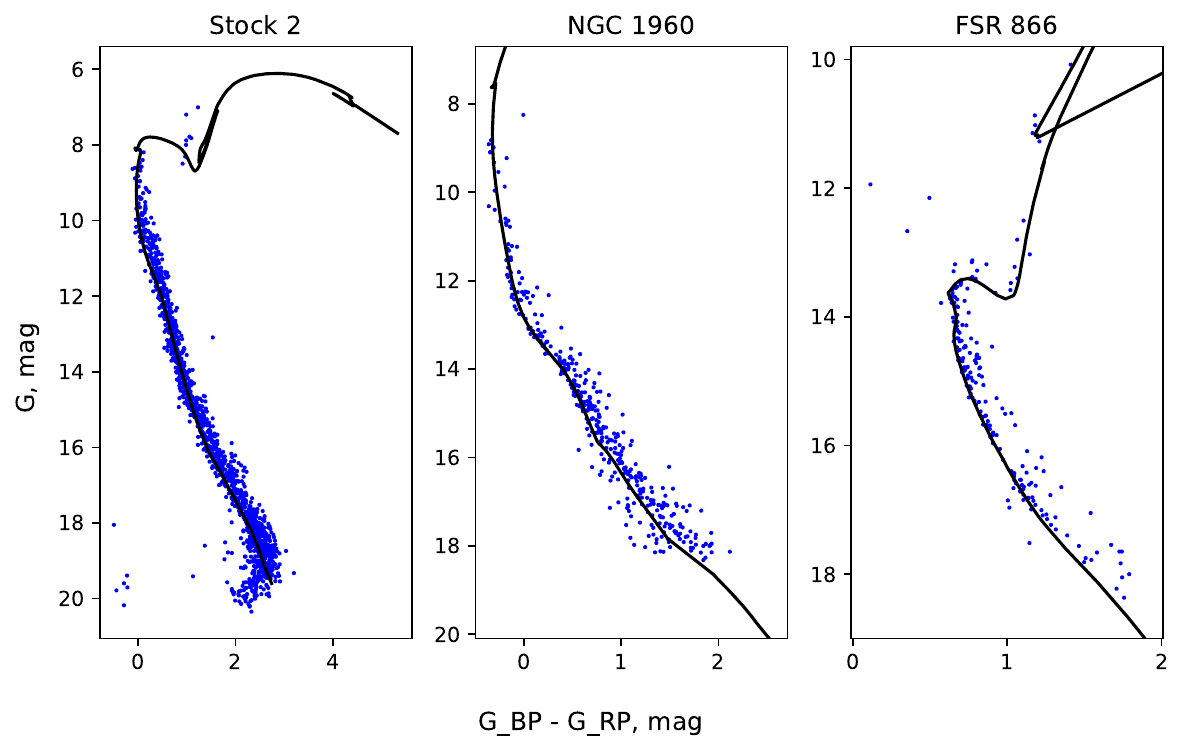}
\caption{Color-Magnitude Diagram (CMD) for Stock 2, NGC 1960 and FSR 866 open cluster based on stars with the membership probability $P \ge 0.95$ for Stock 2, NGC 1960 and $P \ge 0.985$ for FSR 866. Colors are preliminarily dereddened. Parameters are as follows ($DM_{\rm G}$ is the distance modulus in the Gaia filter G): Stock 2: $\log{(\mathrm{Age/yr})} =$ 8.5, $E(B-V) = 0.336$, $DM_{\rm G} = 9.2$ mag, Distance = 463 pc; NGC 1960: $\log{(\mathrm{Age/yr})} =$ 7.48, $E(B-V) = 0.27$, $DM_{\rm G} = 11.1$ mag, Distance = 1202 pc; FSR866: $\log{(\mathrm{Age/yr})} =$ 9.57, $E(B-V) = 0.05$, $DM_{\rm G} = 10.75$ mag, Distance = 1331 pc.}
\label{fig:cmd}
\end{figure*}

In order to use the photometric method to determine the fraction of binaries in an open cluster, we first estimated distance and age of each cluster. We derived cluster parameters by fitting the main sequences of cluster members with theoretical PARSEC + COLIBRI isochrones \citep{Bressan12}, previously correcting the photometry data for differential reddening using color excesses from 3D dust map Bayestar \citep{Green}. The results of isochrone fitting for open clusters Stock 2, NGC 1960 and FSR 866 are shown in the Figure \ref{fig:cmd}. To query for the individual $E(B-V)$ value of any star, distance and coordinates of the source are needed as input parameters. With distance taken from \citet{BailerJones} we got a color excess $E(B-V)$ estimate for each star from the cluster lists and then transformed it into Gaia's color excess $E(G_{\rm{BP}} - G_{\rm{RP}})$ using the relation:
$E(G_{\rm{BP}} - G_{\rm{RP}}) = 3.1*E({\rm B-V})*0.44898$. The coefficient value 0.44898 is for a G2V star, see for instance \citet{Cardelli} and \citet{ODonnell1994} \footnote{\url{http://stev.oapd.inaf.it/cgi-bin/cmd_3.7}}. We then used [Fe/H] from \citet{Dias} to select the isochrone that we over-imposed onto the cluster main sequence.

\section{PHOTOMETRIC BINARY FRACTION}
\label{sec:bin_frac_photo}

A binary system composed of two main sequence stars is brighter than each of its component and redder than the brightest one and therefore is displaced from the main sequence  in the color-magnitude diagram. The difference in colors and magnitudes depends on the mass ratio $q = M_2/M_1 \le 1$ (where $M_1$ and $M_2$ are the masses of the primary and secondary components). In the particular case of  equal mass system ($q = 1$), this results in ~0.75 magnitude shift above the MS. When $q \rightarrow 0$ position of the binary will approach the MS, since the influence of the fainter component will become negligible. To analyse the positions of the stars on the color-magnitude diagram, it is necessary to determine the distance modulus of the cluster and define the binary stars location depending on mass ratio $q$. 

The fraction of  unresolved binary systems was estimated using the approach described in \citet{Milone12}, \citet{Cordoni18}, and \cite{Cordoni23}. First, we used pre-calculated theoretical PARSEC + COLIBRI isochrones \citep{Bressan12} to define the area on color - magnitude diagram where both single and binary cluster stars with $G_{\mathrm{RP}}^{low} \le G_{\mathrm{RP}} \le G_{\mathrm{RP}}^{high}$ lie. We select the isochrone of the age and metallicity, as it was defined in Section \ref{sec:photDist} and calculate new isochrones for the binary star fraction with a given ratio $q$. Then we divided the mentioned area in the colour-magnitude diagram into two parts: the one with predominance of single MS stars and the second, above the MS, which contains candidates binary stars. In order to distinguish them we use a threshold value of $\tilde{q}$. \citet{Cordoni23} used $\tilde{q}$ = 0.6 or 0.7, depending on the cluster. It seems reasonable to not use too small values of q due to possible phothometric errors, so we made calculations for a set of $\tilde{q} = 0.5, 0.6, 0.7, 0.8, 0.9$. Those stars, whose position in the CMD corresponds to the position of binary systems with higher or equal $\tilde{q}$ values, are supposed to be  binary star candidates.
As the cluster`s MS, in order to avoid the overlap of different sequences, we used the portion of the sequence below the main sequence turn off point (see Sec. \ref{sec:photDist}). The pairs of values ($G_{\mathrm{RP}}^{low}, G_{\mathrm{RP}}^{high}$) for clusters turned out to be ($14^m.3$, $22^m.9$) for FSR 866, ($8^m.5$, $21^m.8$) for NGC 1960, and ($9^m.2$, $20^m.8$) for Stock 2.

After that, we built a "box" around the segment of MS under investigation, within which we studied the stars of the cluster in the CMD to determine the binary fraction. To set its left and right borders, we calculated a color standard deviation $\sigma_{color}$ on the basis of intrinsic color errors and the ones from Bayestar \citep{Green} in the following way: (1) to assess the error associated with \citet{Green}, we used data on color excess for stars corresponding to distances from the Bailes-Jones catalog \citep{BailerJones}. Subsequently, for each star, distances were adjusted by adding and subtracting the error values. The difference between the measured \citet{Green} color excess and the corresponding value given the distance error was determined for each star. We selected the maximum value between these two quantities for each star. The average was then computed for all stars with a membership probability exceeding a specified threshold, denoted as $\tilde{p}$, which varies across different clusters. (2) The Gaia color error for each star was calculated according to Equation~12 in \citet{Hallakoun} using BP and RP Gaia flux and the corresponding flux errors. The average error was computed for all stars with a membership probability exceeding the mentioned above threshold, $\tilde{p}$. (3) Finally, the overall error, denoted as $\sigma_{color}$, was obtained by taking the square root of the sum of squares of the \citep{Green} error and the Gaia color error. This parameter, $\sigma_{color}$, represents the measure of the total error and is then used in any further analysis and interpretation of the results.

To set the borders of the "box", we shifted the MS by four times $\sigma_{color}$ to the blue side in order to obtain the left border and, for the right one, we displaced the line of $q = 1$ by four color errors $\sigma_{color}$ to the red side from its original position. The top and bottom of the box are simply the line of binaries with $q$ in all range $0 \le q \le 1$ (calculated with step $\Delta q = 0.1$ and interpolated with splines) for which magnitude of the primary stellar component equals to $G_{\mathrm{RP}}^{high}$ and $G_{\mathrm{RP}}^{low}$, respectively.

All the stars inside this "box" are considered to belong to the cluster. Red-ward of  $q= \tilde{q}$ there are mostly binary MS stars only, and therefore we marked all the members from this area as binaries. The line corresponding to $q = \tilde{q} $ was calculated using the photometric information of the MS stars from the fitted isochrone. 

\begin{figure*}[h]
\plotone{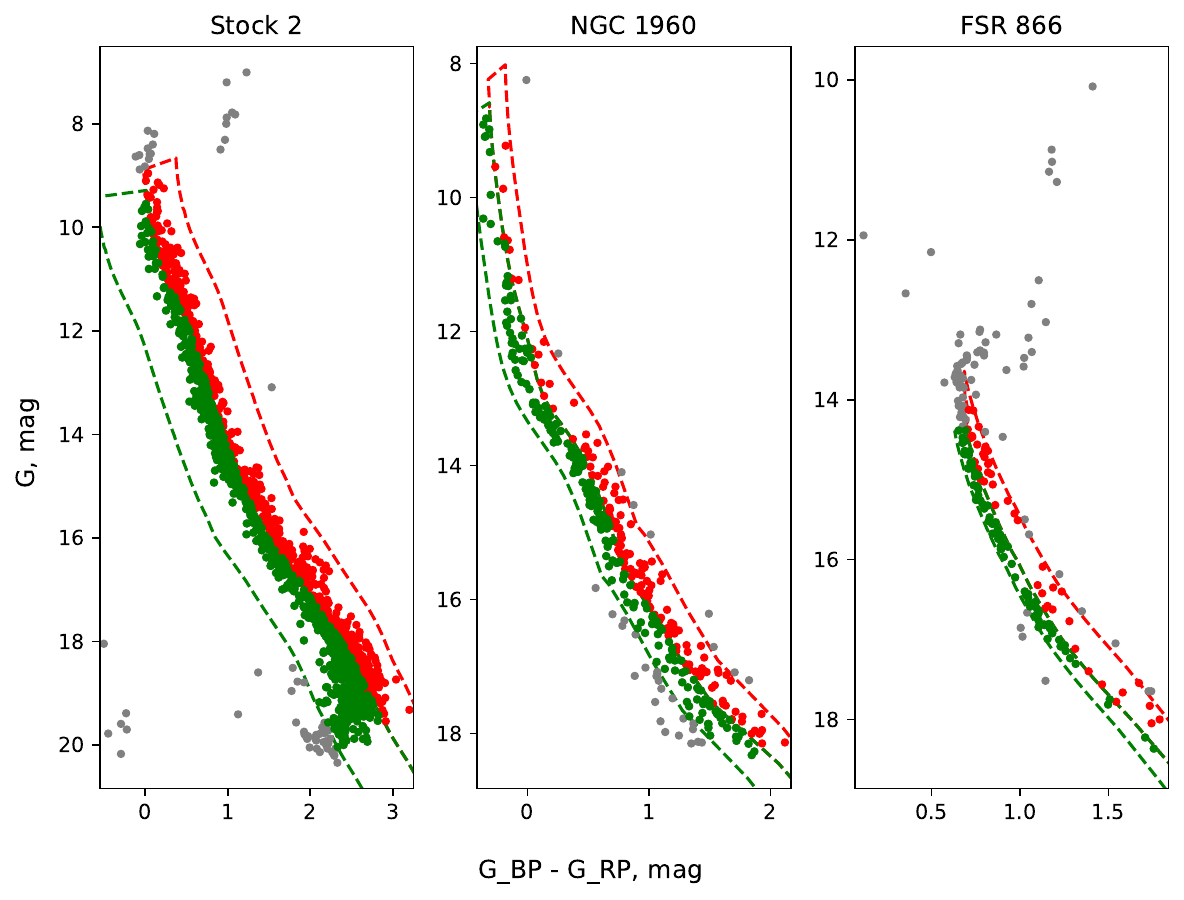}
\caption{Area on the Color-Magnitude Diagram (CMD) for Stock 2, NGC 1960 and FSR 866 open cluster based on stars with the membership probability $P > 0.95$, $P > 0.95$ and $P > 0.985$, respectively, where binary fraction were investigated. The dashed border indicates the boundaries of the study area. For more details on its determination see Sec. \ref{sec:bin_frac_photo}. Grey dots are stars beyond the area. The green part is the area below the line $q = 0.6$, while the red one is the area above the line $q = 0.6$, which is considered to contain binary MS stars. Colors are preliminarily dereddened.}
\label{fig:cmd_box}
\end{figure*}

Figure \ref{fig:cmd_box} shows the investigated area for clusters Stock 2, NGC 1960 and FSR 866 for the case $\tilde{q} = 0.6$. Stars in the region $q < 0.6$ are marked in green while the ones in the binary area $q \ge 0.6$ in red color. 

Binary fractions were the simply calculated as the number of stars within the $q \ge \tilde{q}$ area, $N_{q \ge \tilde{q}}$ divided by total number of stars in the investigated "box" in CMD, $N_{\text{total}}$:

\begin{equation}
    \label{eq:binfrac}
    f_{\text{bin}} = \frac{N_{q \ge \tilde{q}}}{N_{\text{total}}}
\end{equation}

\begin{figure}[h]
\epsscale{1.25}
\plotone{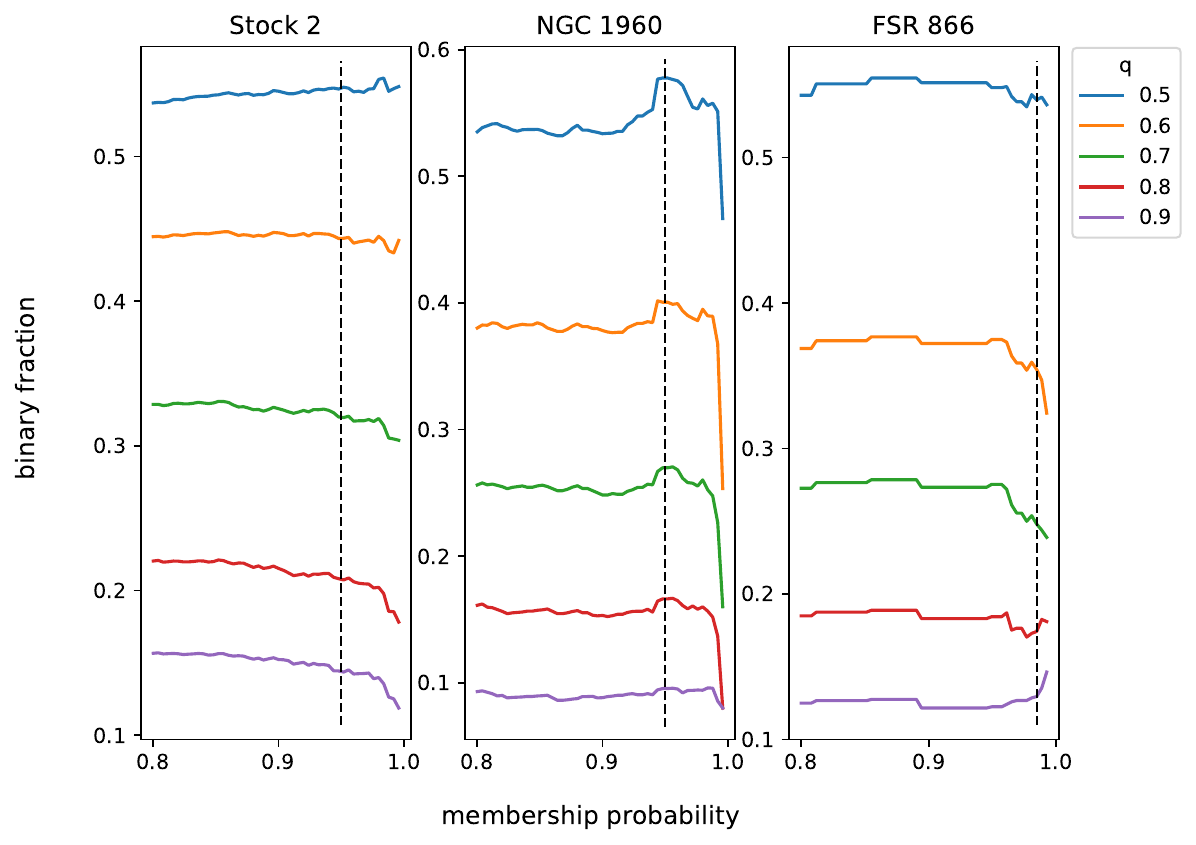}
\caption{Dependency of binary fraction on members selection for different values of the mass ratio q, determined by membership probability threshold, for Stock 2, NGC 1960 and FSR 866 open clusters. The plot covers a threshold range from 0.8 to 1.0. Black dashed line shows the membership probability threshold (0.95 for Stock 2 and NGC 1960 and 0.985 for FSR 866.)}
\label{fig:bin_prob}
\end{figure}

The same procedure was repeated in the wide range of possible membership probability thresholds from 0.8 to 1, in order to track the dependency of the binary fraction on members selection. These dependencies are shown in Figure \ref{fig:bin_prob}. The dependence can be described as an approximately constant value (plateau) or as small variations around the mean and a sharp change in the fraction of binary when the threshold probability approaches unity, when only a few stars with the highest membership confidence are considered as members of the cluster.

The results of the binary fractions estimations for different $\tilde{q}$ is presented in the Table~\ref{tab:BF}.

\begin{table}[h]
\caption{Fraction of binaries depending on choosing mass ratio threshold $\tilde{q}$.}
\label{tab:BF}
\centering
\begin{tabular}{lccccc}
\hline
OC & \multicolumn{5}{c}{Binary fraction} \\
 & $\tilde{q}$ = 0.5 & $\tilde{q}$ = 0.6 & $\tilde{q}$ = 0.7 & $\tilde{q}$ = 0.8 & $\tilde{q}$ = 0.9 \\
\hline
FSR 866 & 0.54 & 0.35 & 0.25 & 0.17 & 0.13\\
NGC 1960 & 0.58 & 0.4 & 0.27 & 0.17 & 0.1 \\
Stock 2 & 0.55 & 0.44 & 0.32 & 0.21 & 0.14 \\
\hline
\end{tabular}
\end{table}

\section{SPECTROSCOPIC OBSERVATIONS AND DATA REDUCTION}
\label{sec:ObsPhot}

We exploited spectroscopic data obtained during a run in the period 2019 - 2021 carried out at the Asiago Astrophysical Observatory, Asiago, Italy.
We used the 122cm Galileo telescope equipped with the Boller\&Chivens + CCD spectrograph with the 1200 ln/mm grating and range $\lambda = 3820-5035 \text{\AA}$ (spectral resolution R $\approx 4000$) and 1.8m Copernico telescope that we used for Echelle spectra ($\lambda = 3470-7360 \text{\AA}$, \textbf{R $\approx 20000$}).

The CCD frames underwent reduction using the well-established IRAF packages. Flat fields were consistently obtained for each observing run, and calibration spectra were recorded using a stationary Fe/Ar lamp before observing the target star.

Radial velocities were derived from the observational data using of Fourier cross-correlation technique from the IRAF software package. As a template we used a synthetic spectra. We first determined for each of the target star spectral class according to \citet{Gray2009}, then select atmospheric parameters for the inferred spectral types from \citet{Landolt1982} and then according to the selected parameters we chose a synthetic spectra based on Kurucz's codes from \citet{Munari2005}. Together with the measure of  radial velocities of the target stars, we also provided estimates of their spectral types. They are listed in a Table \ref{tab:spec_class} which is divided into three blocks with member-stars of each of the three studied open clusters.

\begin{table}[h]
\centering
\caption{List of member stars for OCs under investigation with spectral classes pointed.}
\label{tab:spec_class}
\begin{tabular}{cc|cc}
\hline
Star No. & SP & Star No. & SP \\ [0.5ex] 
\hline\hline

\multicolumn{4}{c}{FSR 866} \\
\hline
s1 & K2 III &  s8  & K0 V     \\
s2 & K0 III &  s10 & K0 V     \\
s3 & K0 III &  s11 & A7 - F0  \\
s5 & K0 III &  s12 & F8       \\
s6 & F2     &  s22 & G0       \\
s7 & F5     &  s32 & K0       \\
\hline
\multicolumn{4}{c}{NGC 1960} \\
\hline
s4 & B3       &  s12 & B5         \\
s5 & B1       &  s13 & B7         \\
s6 & B3       &  s14 & B5         \\
s7 & B1e$_3$  &  s15 & B8 	      \\
s8 & B1 V     &  s17 & B7 - B8    \\
s9 & B5e$_2$  &  s18 & B6         \\
s11 & B3      &  s19 & B7         \\   
\hline
\multicolumn{4}{c}{Stock 2} \\
\hline
s2 & G8 III           & s26 & A1 \\
s3 & G8 III - K0 III  & s27 & A0 \\  
s7 & A1               & s28 & A1 \\ 
s8 & A2               & s31 & A2 \\ 
s9 & G8 III           & s32 & A2 \\ 
s10 & A2              & s33 & A1 \\ 
s11 & A1              & s36 & A2 \\ 
s12 & G8 III          & s41 & A2 \\ 
s13 & A1              & s42 & A1 \\ 
s15 & A1              & s43 & A1 \\          
s16 & A1 - A2         & s44 & A1 \\  
s17 & A1              & s45 & A1 \\  
s18 & A2              & s46 & A2 \\
s19 & A1              & s47 & A4 \\
s21 & A0              & s49 & A3 \\
s22 & A2              & s50 & A2 \\
s23 & A1              & s51 & A2 \\
s24 & A2              & s68 & A3 \\
s25 & A2              & & \\
\hline

\end{tabular}
\end{table}

\section{BINARY FRACTION ACCORDING TO SPECTROSCOPY AND MEAN RADIAL VELOCITY}
\label{sec:bin_frac_velo}

\begin{figure}
\plotone{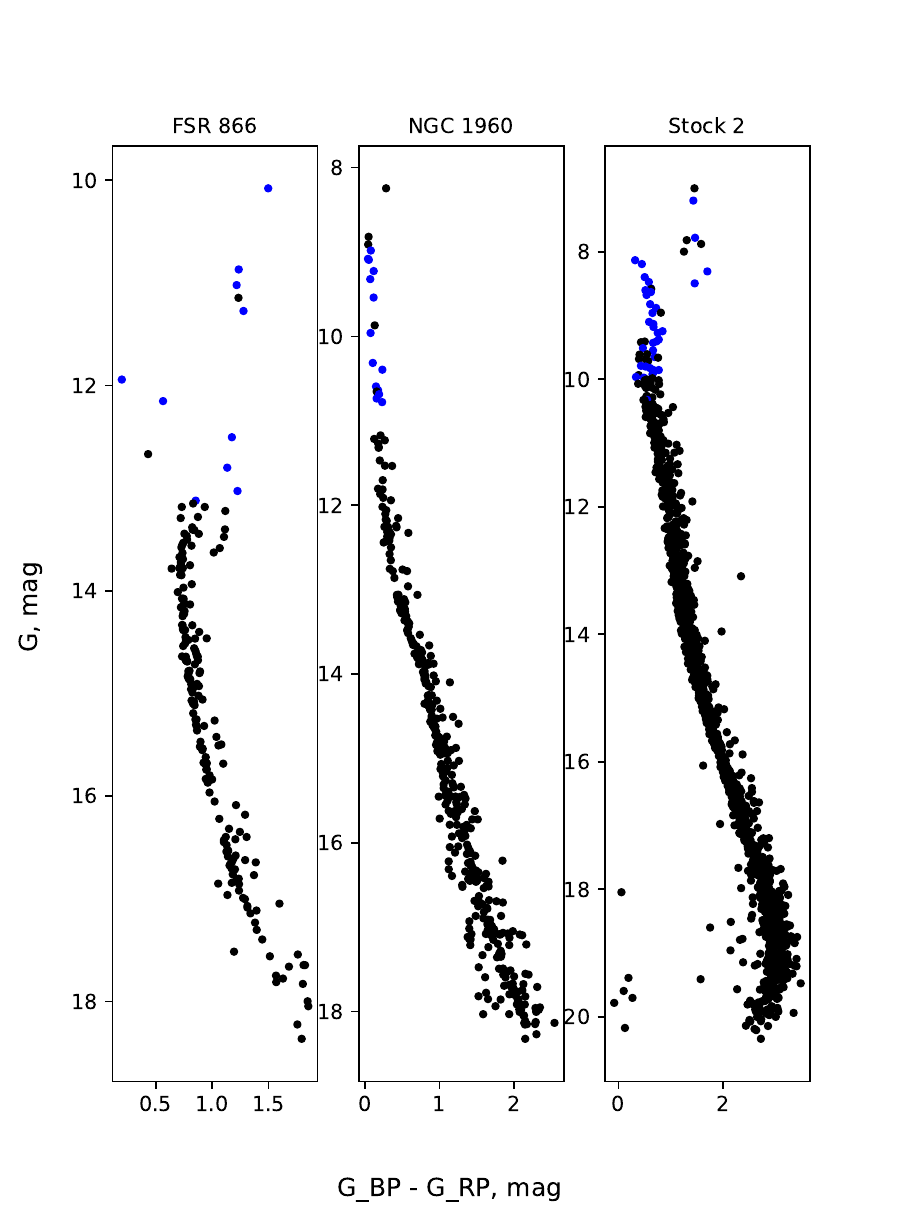}
\caption{Color-Magnitude Diagram (CMD) for Stock 2, NGC 1960, and FSR 866 open cluster, only stars with the membership probability P $\ge$ 0.95 for Stock 2, NGC
1960 and  P $\ge$  0.985 for FSR 866 are shown; blue color indicates stars with spectroscopic measurements.}
\label{fig:rv_cmd}
\end{figure}

The Figure~\ref{fig:rv_cmd} shows the CMD of the clusters with stars for which we have the spectroscopic measurements depicted with blue color.

\begin{figure*}
\plotone{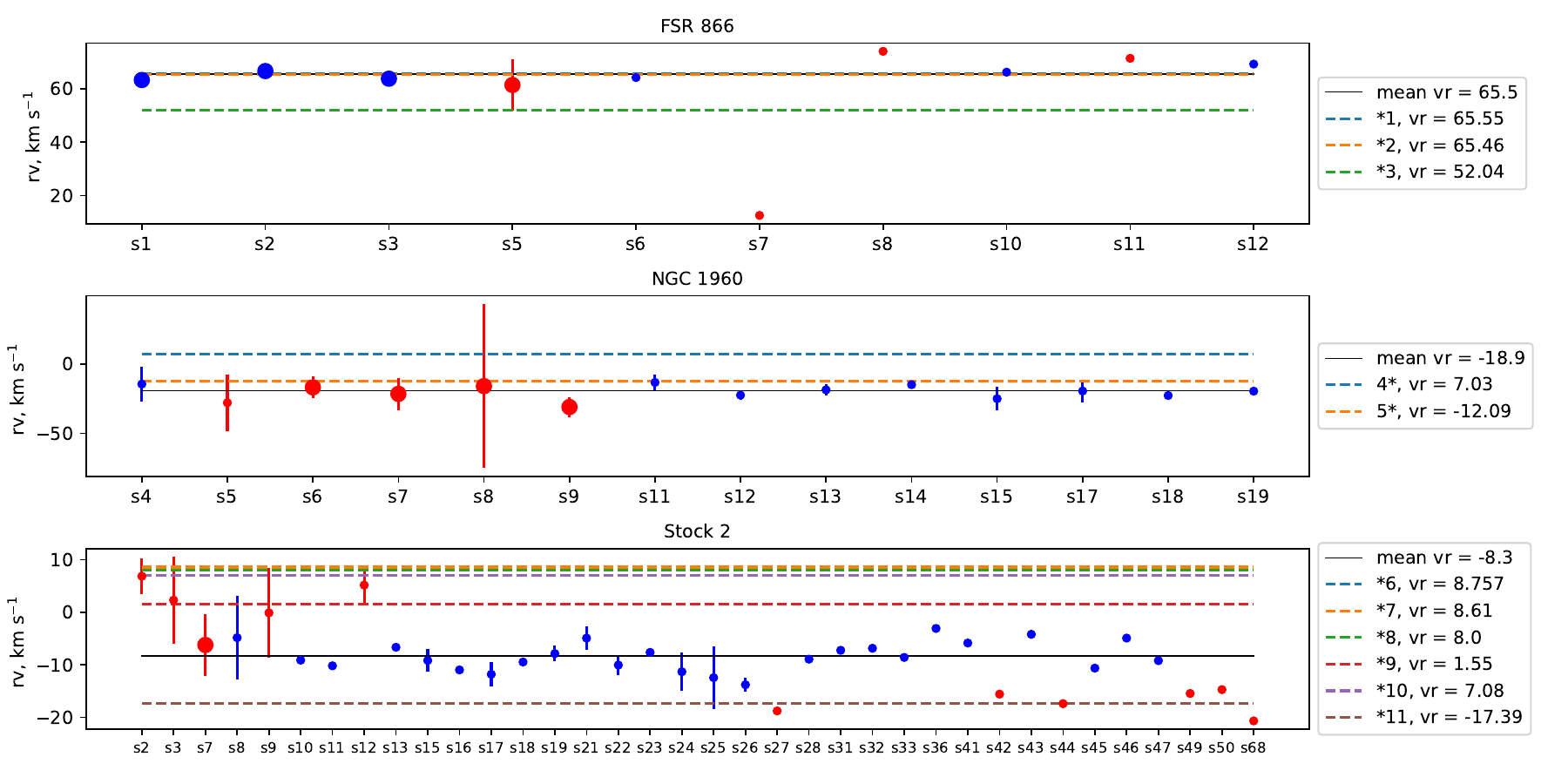}
\caption{Radial velocity measurements for the clusters. Single stars are depicted with blue color and double stars are with red. Large circles indicates the stars with more than 3 observations. Black line is a mean radial velocity, other lines demonstrate the estimates from the literature that were mentioned in the Table~\ref{tab:Literature}, where *1 - \citet{Dias}, *2 - \citet{Tarricq2021}, *3 - \citet{Zhong2020}, 4* - \citet{Tarricq2021}, 5* - \citet{Zhong2020}, *6 - \citep{Dias}, *7 - \citet{Tarricq2021}, *8 - \citet{AlonsoSantiago2021}, *9 - \citet{Zhong2020}, *10 - \citet{Carrera2019}, *11 - \citet{Mermilliod2008}}
\label{fig:rv}
\end{figure*}

The results of the radial velocities extraction are presented in Tables \ref{tab:FSR866}, \ref{tab:NGC1960}, and \ref{tab:Stock2} for FSR 866, NGC 1960, and Stock 2, respectively, and in  Figure~\ref{fig:rv}. In order to select binary stars, we used the following criteria \citep{ngc225}:

\begin{enumerate}
\item[a)] If a star had more than 3 observations, we used the Pearson's chi-squared test with 95$\%$ significance level.
\item[b)] If a star had only 3 or less observations, we calculated mean radial velocity $(V_r)$ and standard deviation $(\sigma V_r)$ and then compared them with the cluster mean radial velocity $(V_{r,\mathrm{Cl}})$ and its error $(\sigma V_{r,\mathrm{Cl}})$. If $V_r$ differs from $V_{r,\mathrm{Cl}}$ by $3 \times \sigma V_{r,\mathrm{Cl}}$ or $\sigma V_r > 3 \times \sigma V_{r,\mathrm{Cl}}$, the star was assumed to be a candidate binary.
\end{enumerate}

The mean cluster radial velocity and its error were calculated using  an iterative process. We used only {\it bona fide} single stars. At the first step, we have taken only stars marked as single by criterion (a). Then stars marked as single by criterion (b) were added and values of $V_{r,\mathrm{Cl}}$ and $\sigma V_{r,\mathrm{Cl}}$ were re-calculated. The iterative process were repeated until convergence, when no additional single stars are left. The only exception to this algorithm was Stock 2 cluster, because for this cluster at the 1st step criterion (a) gives no single stars with greater than 3 observations for the iteration process to be continued. It is connected with the high value of the initial estimation of $\sigma V_{r,\mathrm{Cl}}$, so no stars are marked as single initially. For Stock 2 we used only criterion (b) for the iterative process, starting the zero iteration with the stars which lie within only one $\sigma V_{r,\mathrm{Cl}}$ and continuing iterations as usual. 

The resulting values of mean clusters radial velocities and their errors, calculated on the basis of single stars only, are as follows: 65.5 $\pm$ 1.0 km s$^{-1}$ for FSR 866, -18.9 $\pm$ 1.4 km s$^{-1}$ for NGC 1960 and -8.3 $\pm$ 0.5 km s$^{-1}$ for Stock 2.

Following the above criteria, we estimated the binary status of stars with measured radial velocity. The final result for each star with measured radial velocity is presented in the Tables \ref{tab:FSR866}, \ref{tab:NGC1960} and \ref{tab:Stock2}.

To estimate a binary fraction for a cluster, one needs to estimate the potential binary nature for \textit{each} star in some magnitude range. We sorted stars with measured radial velocities (and so with its binary status estimated) in descending order of magnitude and checked whether there are more member-stars (i.e. stars with membership probability higher than a chosen threshold) in this magnitude range which do not have their binary status assessed, to build up a complete sample of cluster member stars in this magnitude range. The resulting binary fraction estimations are as follows:

\begin{enumerate}
\item[] \textit{FSR 866}. We have 10 member-stars with measured radial velocities, 4 of which are considered to be binaries. In addition, there are two more stars in the considered magnitude range without a corresponding estimations, so they could be either binary or single. The binary fraction for FSR 866 lies between 4/12 (0.33) and 6/12 (0.50), hence 0.42 on average. 
\item[] \textit{NGC 1960}. We have 14 member-stars with measured radial velocities, 5 of which are considered to be binaries. In addition, there are two more stars in the magnitude range under investigation without a corresponding estimations, so they could be either binary or single. The binary fraction for NGC 1960 lies between 5/16 (0.31) and 7/16 (0.44), hence 0.38 on average.
\item[] \textit{Stock 2}. For this cluster we did not take into account the star s68, because it is much fainter than the rest and brings together a lot of stars without radial velocity estimation within the resulting range. Excluding s68 star, we have 35 member-stars with measured radial velocities, 10 of which are considered to be binaries. In addition, there are 14 more stars in the considered magnitude range without a corresponding estimate, so they could be either binary or single. The binary fraction for Stock 2 lies between 10/49 (0.20) and 24/49 (0.49), hence 0.35 on average.
\end{enumerate}

\begin{table}[h]
\caption{Binary fractions (BF) obtained from radial velocities (Sec. \ref{sec:bin_frac_velo}), compared to binary fraction obtained with from photometry (Sec. \ref{sec:bin_frac_photo}) for the case $\tilde{q}$ = 0.6. The last column indicates the photometric binary fraction}
\label{tab:binarity_comp}
\centering
\begin{tabular}{lccc}
\hline
OC & BF range (RV) & Avg BF (RV) & BF ($\tilde{q}$ = 0.6)\\ [0.5ex] 
\hline
FSR 866 & 0.33 -- 0.50 & 0.42 & 0.35  \\
NGC 1960 & 0.31 -- 0.44 & 0.38 & 0.4 \\
Stock 2 & 0.20 -- 0.49 & 0.35 & 0.44 \\
\hline
\end{tabular}
\end{table}

\begin{table}[h]
\centering
\caption{List of stars with measured radial velocities for FSR 866 with indices in decreasing order of $G_{\rm{mag}}$ (See Sec. \ref{sec:clus_algo}). $N_{obs}$ - number of observations, $V_r$ - mean radial velocity, $\sigma V_r$ - its error, Binarity - binary status based on the (a) and (b) criteria as described in the Sec. \ref{sec:bin_frac_velo}.}
\label{tab:FSR866}
\begin{tabular}{lcccc}
\hline
\multicolumn{5}{c}{FSR866} \\
\hline
\hline
Star & $N_{obs}$ & $V_r$ & $\sigma V_r$ & Binarity \\ [0.5ex] 
\hline
s1 & 5 & 63.267 & 0.872 & single \\
s2 & 5 & 66.622 & 1.821 & single\\
s3 & 4 & 63.767 & 2.113 & single\\
s5 & 4 & 61.413 & 9.542 & binary\\
s6 & 2 & 64.189 & 0.267 & single \\
s7 & 1 & 12.509 & 0.0 & binary  \\
s8 & 1 & 74.019 & 0.0 & binary  \\
s10 & 1 & 66.204 & 0.0 & single  \\
s11 & 1 & 71.406 & 0.0 & binary  \\
s12 & 1 & 69.224 & 0.0 & single \\
\hline

\end{tabular}
\end{table}

\begin{table}[h]
\centering
\caption{Same as Table \ref{tab:FSR866}, but for NGC 1960.}
\label{tab:NGC1960}
\begin{tabular}{lcccc}
\hline
\multicolumn{5}{c}{NGC1960} \\
\hline
\hline
Star & $N_{obs}$ & $V_r$ & $\sigma V_r$ & Binarity \\ [0.5ex] 
\hline
s4 & 3 & -14.389 & 12.497 & single   \\
s5 & 3 & -27.911 & 20.634 & binary   \\
s6 & 4 & -16.782 & 7.976 & binary   \\
s7 & 4 & -21.469 & 11.583 & binary   \\
s8 & 5 & -15.867 & 59.392 & binary  \\
s9 & 4 & -30.987 & 7.378 & binary \\
s11 & 3 & -13.194 & 5.808 & single   \\
s12 & 3 & -22.391 & 3.72 & single    \\
s13 & 2 & -18.548 & 3.791 & single   \\
s14 & 2 & -14.833 & 0.863 & single   \\
s15 & 2 & -24.959 & 8.597 & single   \\
s17 & 2 & -19.387 & 8.172 & single   \\
s18 & 2 & -22.659 & 1.808 & single   \\
s19 & 2 & -19.527 & 1.882 & single   \\
\hline

\end{tabular}
\end{table}

\begin{table}[h]
\centering
\caption{Same as Table \ref{tab:FSR866}, but for Stock 2.}
\label{tab:Stock2}
\begin{tabular}{lcccc}
\hline
\multicolumn{5}{c}{Stock 2} \\
\hline
\hline
Star & $N_{obs}$ & $V_r$ & $\sigma V_r$ & Binarity \\ [0.5ex] 
\hline
s2 & 3 & 6.884 & 3.399 & binary     \\
s3 & 3 & 2.334 & 8.243 & binary     \\
s7 & 4 & -6.204 & 5.945 & binary    \\
s8 & 2 & -4.824 & 7.904 & single    \\
s9 & 3 & -0.087 & 8.605 & binary    \\
s10 & 1 & -9.101 & 0.0 & single     \\
s11 & 2 & -10.178 & 0.708 & single  \\
s12 & 3 & 5.192 & 3.545 & binary    \\
s13 & 2 & -6.646 & 0.427 & single   \\
s15 & 2 & -9.159 & 2.117 & single   \\
s16 & 1 & -10.956 & 0.0 & single    \\
s17 & 2 & -11.782 & 2.249 & single  \\
s18 & 1 & -9.451 & 0.0 & single     \\
s19 & 3 & -7.854 & 1.52 & single    \\
s21 & 2 & -4.9 & 2.268 & single     \\
s22 & 2 & -10.032 & 1.923 & single  \\
s23 & 1 & -7.616 & 0.0 & single     \\
s24 & 2 & -11.311 & 3.675 & single  \\
s25 & 2 & -12.433 & 5.918 & single  \\
s26 & 2 & -13.781 & 1.291 & single  \\
s27 & 1 & -18.762 & 0.0 & binary    \\
s28 & 1 & -8.907 & 0.0 & single     \\
s31 & 1 & -7.224 & 0.0 & single     \\
s32 & 1 & -6.845 & 0.0 & single     \\
s33 & 1 & -8.574 & 0.0 & single     \\
s36 & 1 & -3.056 & 0.0 & single     \\
s41 & 1 & -5.835 & 0.0 & single     \\
s42 & 1 & -15.563 & 0.0 & binary    \\
s43 & 1 & -4.179 & 0.0 & single     \\
s44 & 1 & -17.376 & 0.0 & binary    \\
s45 & 1 & -10.615 & 0.0 & single    \\
s46 & 1 & -4.9 & 0.0 & single       \\
s47 & 1 & -9.19 & 0.0 & single      \\
s49 & 1 & -15.446 & 0.0 & binary    \\
s50 & 1 & -14.704 & 0.0 & binary    \\
s68 & 1 & -20.665 & 0.0 & binary    \\
\hline

\end{tabular}
\end{table}

\section{SUMMARY AND  CONCLUSIONS}
In this work we performed a photometric and spectroscopic study of three open star clusters: FSR 866, NGC 1960 and Stock 2, with the goal of estimating the fraction of binaries among their member stars. We used Gaia DR3 photometry and spectroscopy for pre-selected bright member-stars of these clusters, conducted by us at Asiago Astrophysical Observatory, Asiago, Italy over the years 2019 - 2021. We obtained a list of probable cluster members by applying the DBSCAN clustering algorithm in 3-dimensional space of proper motion and parallaxes exploring a wide range of algorithm parameters. 
 
 To obtain clusters' parameters (age and distance) we first used Bayestar 3D dust map \citep{Green} to calculate individual reddening for each cluster member and shifted them according to  this value. The mean color excesses, obtained on the basis of individual members reddening are as follows:  $E(B-V)=0.34$ for Stock 2, $E(B-V)=0.27$ for NGC 1960 and $E(B-V)=0.05$ for FSR 866. After that we fitted theoretical PARSEC + COLIBRI isochrones \citep{Bressan12} onto reddening-corrected $(G, BP-RP)$ color-magnitude diagram and got the following age and distance estimates: $\log{(\mathrm{Age/yr})}$ = 8.50 and distance = 463 pc for Stock 2, $\log{(\mathrm{Age/yr})}$ = 7.48 and distance = 1202 pc for NGC 1960, and $\log{(\mathrm{Age/yr})}$ = 9.57 and distance = 1330 pc for FSR 866.

 We employed two different methods to binary fraction estimation: photometric and spectroscopic. While studying spectra of member stars, we also derived their spectral classification, which is listed in Table \ref{tab:spec_class}.

 As for photometry, we followed the approach described by \citet{Milone12}, \citet{Cordoni18} and \citet{Cordoni23}). We investigated "boxes" on CMD of each cluster with the boundaries determined by certain magnitude range, main sequence position and color scatter of cluster members (for details see Sec. \ref{sec:bin_frac_photo}).  
 The resulting binary fractions for three clusters for different threshold parameter of mass ratio $\tilde{q}$ are reported in the Table \ref{tab:BF} and for the case $\tilde{q} = 0.6$ are: 0.35 for FSR 866, 0.4 for NGC 1960 and 0.44 for Stock 2. All values are within the typical range for previously investigated OCs.

 As for the spectroscopic, we used radial velocities of cluster members, obtained by us from our spectroscopic observations (see Tables \ref{tab:Stock2}, \ref{tab:NGC1960} and \ref{tab:FSR866}).  The resulting values of mean clusters radial velocities and their errors, calculated on the basis of single stars only, are as follows: 65.5 $\pm$ 1.0 km s$^{-1}$ for FSR 866, -18.9 $\pm$ 1.4 km s$^{-1}$ for NGC 1960 and -8.3 $\pm$ 0.5 km s$^{-1}$ for Stock 2.  
 The resulting values for the binary fraction in the three OCs are as follows (see Table \ref{tab:binarity_comp}): for FSR 866 it lie in the range  4/12 (0.33) - 6/12 (0.50), for NGC 1960  5/16 (0.31) - 7/16 (0.44), and for Stock 2 10/49 (0.20) - 24/49 (0.49). Again, all ranges are quite typical for OCs as general.

 The two approaches for estimating the binary are complementary: photometry method provides insights into the fainter part of the main sequence, while radial velocities mostly allow one to check the binary status of more massive and bright cluster members. Of course, both approaches have their limitations in revealing binary systems. The main limitations are the uncertainty due to photometric errors and dependence on threshold parameter $\tilde{q}$ for the photometric approach, and the need for a large number of repeated observations for the spectroscopy method. As mentioned, the binary fraction depends on many factors, such as age, mass, and density of the cluster. The three investigated clusters differ in all these parameters, making them difficult to compare. Since the photometry approach takes into account more stars, it can be considered more reflective of the average number of binary systems in cluster. Then it can be noticed that FSR 866, being the oldest of the three clusters, has the lowest fraction of binaries according to the mentioned approach, while binary fraction of two younger clusters, NGC 1960 and Stock 2, do not differ significantly.

 The complete lists of member stars with marks of those for which radial velocities have been measured are available for all three clusters as additional materials to the article.

\begin{acknowledgments}
The reported study was funded by RFBR according to the research project number 20-32-90124. We are grateful to U. Munari and P. Ochner for securing part of the observational data and for their valuable advice on the data processing. This work has made use of data from the European Space Agency (ESA) mission
{\it Gaia} (\url{https://www.cosmos.esa.int/gaia}), processed by the {\it Gaia}
Data Processing and Analysis Consortium (DPAC,
\url{https://www.cosmos.esa.int/web/gaia/dpac/consortium}). Funding for the DPAC
has been provided by national institutions, in particular the institutions
participating in the {\it Gaia} Multilateral Agreement.  
\end{acknowledgments}

%

\vspace{5mm}
\facilities{Asiago Astrophysical Observatory}


\software{DBSCAN \citep{dbscan},  
          \textsc{scikit-learn} \citep{scikit-learn}, 
          }



\bibliography{bib}{}
\bibliographystyle{aasjournal}



\end{document}